\documentstyle[aps,prl,epsfig,multicol,floats]{revtex}

\begin{document}
\draft

\twocolumn[\hsize\textwidth\columnwidth\hsize\csname@twocolumnfalse%
\endcsname

\title{Brownian Motion in Gravitationally-Interacting Systems}

\author{Pinaki Chatterjee, Lars Hernquist \&
         Abraham Loeb}
\address{Harvard-Smithsonian Center for Astrophysics, 60 Garden Street,
Cambridge, MA 02138, USA}
\address{pchatterjee@cfa.harvard.edu,
lars@cfa.harvard.edu, aloeb@cfa.harvard.edu}

\date{\today}

\maketitle

\begin{abstract}

We derive a model that describes the motion of a Brownian particle in a
system which is dominated by gravitational forces. An example of such a
system is a massive black hole immersed in a cluster of stars.  We compute
the dispersion in the position and velocity of such a black hole, and
examine whether it achieves a state of equipartition of kinetic energy with
the stars.  This problem has been considered before only for stellar
systems with an isothermal Maxwellian distribution of velocities; here we
study other examples and confirm our calculations with N-body simulations.
In certain cases, depending on the stellar distribution function, the black
hole can acquire a steady state kinetic energy which is very far from
equipartition relative to the stars.

\end{abstract}


\vskip2pc]
\flushbottom
\bibliographystyle{unsrt}

Brownian motion in stellar systems was studied by Chandrasekhar
\cite{chandra43a}, who recognized that the force acting on
an object in a stellar system consists of two independent
contributions: one part, which originates from the ``smoothed-out'' 
average distribution of matter in the stellar system, varies slowly 
with position and time; the second part, which arises from
discrete encounters with individual stars, fluctuates much more
rapidly. The smooth force itself consists of the restoring force
arising from the potential of the aggregate distribution of stars, and
the dissipative force of dynamical friction which causes the object to
decelerate as it moves through the stellar background
\cite{chandra43b}. This is similar to the Langevin
model of Brownian motion, which describes the irregular motions of
dust grains immersed in a gas of light molecules: the Brownian
particle experiences a frictional force proportional to its velocity,
and a random, rapidly fluctuating force owing to the large
rate of collisions it suffers with the gas molecules in its
neighborhood. 

In a stellar system, the analog of the gas molecules is the stars, and
the Brownian particle corresponds to any object which is much more
massive than the stars and moves much more slowly than they do. An
astrophysically relevant example is a massive black hole at the center
of a dense stellar system such as a globular cluster or 
the nucleus of a galaxy. We
would like to extend the Langevin method of analysis to such a
problem. 

Consider, therefore, a black hole of mass $m$ in a cluster of 
stars of total mass 
$M$ which we take to be described by the spherically symmetric density
and potential profiles $\rho(r)$ and $\Phi(r)$, respectively, where 
$\mathbf{r}$ is the radial position vector from the center of the 
stellar system, which is taken as the origin. The phase space
distribution function, which in general will depend both on 
position $\mathbf{r}$, and velocity $\mathbf{v}$, is defined such that 
$f({\bf r}, {\bf v})\, d^3{\bf r}\,d^3{\bf v}$ is the mass in stars in 
the phase space volume $d^3{\bf r}\, d^3{\bf v}$. We 
assume that in a spherically symmetric stellar model, $f$ is a
function only of the energy per unit mass $E$, where $E = \frac{1}{2}
v^2 + \Phi (r)$.

A black hole in this stellar system is subject to three forces. The
first is the restoring force of the stellar potential, ${\bf F} = - m 
\nabla \Phi (r)$, where ${\bf r}$ is the position vector of the black 
hole. For a stellar system with a non-cuspy core, this reduces to the
form ${\bf F} = - k {\bf r}$, where 
$k$ is independent of $r$ for small values of $r$.

The second force on the black hole is the dissipative
force of dynamical friction which causes it to decelerate as it moves
through the stellar background. We use for this the well-known Chandrasekhar 
formula 
(\cite{chandra43b}, \cite{BT})
${\bf F} = - \beta {\bf v}$, where 
\begin{equation}
\beta=16 {\pi}^2 {\rm ln} \Lambda\, G^2 m (m+m_{\star}) 
\frac{\int_{0}^{v}f(r,u) u^2 du}{v^3}.
\end{equation}
In the above, ${\bf v}$ is the velocity of the black hole, $m_{\star}$
is the mass of each star (in the following, we take all stars to have
equal masses, for simplicity), and ${\rm ln} \Lambda$ is the ``Coulomb
logarithm'' \cite{BT}.
Since the black hole remains close to the origin and moves very 
slowly compared with the stars, we may replace $f(r,u)$ in the 
integral by $f(r,0)$, and evaluate it at $r \rightarrow 0$ to obtain 
\cite{BT}:
\begin{equation}
\beta \rightarrow (16 \pi^2/3) {\rm ln} \Lambda\, G^2 m (m+m_{\star}) 
f(0,0).
\end{equation} 

The third force on
the black hole fluctuates on a time-scale which is extremely short
compared to the above two forces \cite{CHL}, and arises from random 
discrete encounters between the black hole and the stars. Denoting 
this stochastic force as ${\bf F}(t)$, we can write the equation of
motion of the black hole as 
\begin{equation} 
m \ddot{{\bf r}}(t) + \beta \dot{{\bf r}}(t) + k {\bf r}(t) = {\bf F}(t),
\end{equation}
which is the equation of motion of a harmonically bound Brownian
particle. 

The spatial components of this linear vector equation are separable 
into equivalent components:
\begin{equation} \label{eqofmotion}
m \ddot{x}(t) + \beta \dot{x}(t) + k x(t) = F_x(t).
\end{equation}
The stochastic force is defined only statistically. Since it is random
and rapidly varying, we expect it to be 
zero on average and uncorrelated with itself at different times: 
\begin{equation} \label{stochdef}
\langle F_x(t) \rangle = 0, \qquad
\langle F_x(t_1)\,F_x(t_2) \rangle = C\, \delta (t_1-t_2),
\end{equation}
where $\delta$ is the Dirac delta function and the angular brackets 
denote an average over an ensemble of ``similarly prepared'' systems 
of stars in each of which the black hole has the same initial position 
and velocity. We will later show how to determine the value of $C$. 

We have shown in detail in Ref. \cite{CHL} how Eqs. (\ref{eqofmotion}) and
(\ref{stochdef}) can be combined with the Fokker-Planck equation to derive
the probability distributions of the black hole's position and velocity in
the stationary state (i.e., when initial transients -- which decay
exponentially fast -- die out and these distributions become
time-independent); these turn out to be independent Gaussian distributions:
\begin{equation} \label{FPsolstx}
W(x)=\sqrt{2 \gamma/\pi C} \,\omega_0 \,m\,\, 
\textrm{exp} \lbrack - (2 \gamma/C) \omega_0^2 m^2 x^2 
\rbrack,
\end{equation}
\begin{equation} \label{FPsolstv}
W(v_x)=\sqrt{2 \gamma/\pi C} \,m\,\, 
\textrm{exp} \lbrack - (2 \gamma/C) m^2 v_x^2 
\rbrack,
\end{equation}
where $\gamma = \beta/2m$ and $\omega_0=\sqrt{k/m}$.
Note that $v_x \equiv \dot{x}$.

It was shown in \cite{BT} (their Eq. [8-66]) that the rate of
change of kinetic energy of the black hole is given by
\begin{eqnarray}
d(mv^2/2)/dt & = & 16 {\pi}^2 {\rm ln} \Lambda\, 
G^2 m m_{\star}
\Biggl\lbrack \int_{v}^{\infty} f(r,u) u du {}\nonumber\\ & & {}
- m/(m_{\star} v) 
\int_{0}^{v} f(r,u) u^2 du \Biggr\rbrack.
\end{eqnarray}
The first term describes the ``heating'' of the black hole by
fluctuations in the
stellar system, and the second term describes the ``cooling'' or
dissipative effect of
dynamical friction. Averaged over the ensemble in the stationary state,
the two terms above should sum to zero. Again, 
setting $f(r,u)$ in integrals of the form $\int_{0}^{v} \ldots f(r,u) du$
to $f(r,0)$, 
we can derive 
\begin{equation} \label{vel}
\langle v^2\rangle= (3 m_{\star}/m) \int_{0}^{\infty} 
f(r,u) u du \bigg/f(r,0) = 3 \langle v_x^2\rangle,
\end{equation}
since the three velocity components are
equivalent and independent of each other. From 
Eqs. (\ref{FPsolstx}) and (\ref{FPsolstv}), 
\begin{equation} \label{xvsquare}
\langle x^2\rangle=C/4 \gamma m^2 \omega_0^2, \qquad
\langle v_x^2\rangle=C/4 \gamma m^2.
\end{equation}
Combining Eqs. (\ref{vel}) and (\ref{xvsquare}), 
we obtain the following expressions for $\langle
x^2\rangle$ and $C$:
\begin{equation} \label{x}
\langle x^2\rangle= (m_{\star}/m \omega_0^2) \int_{0}^
{\infty} f(r,u) u du \bigg/f(r,0),
\end{equation}
\begin{equation} \label{C}
C= 4 \gamma m m_{\star} \int_{0}^
{\infty} f(r,u) u du \bigg/f(r,0).
\end{equation}

We wish to have a measure for how far the black hole is from
equipartition of kinetic energy between itself and the stars
surrounding it in the core of the stellar system. 
Since the square of the stellar velocity dispersion is 
\begin{equation} 
\overline{v_{\star}^2}=\int_{0}^{\infty} u^2 f(r,u)\,4\pi u^2\,du \bigg/  
\int_{0}^{\infty} f(r,u)\,4\pi u^2\,du,
\end{equation}
we obtain 
by using Eq. (\ref{vel}):
\begin{equation} \label{equi}
\langle v^2\rangle /\,\overline{v_{\star}^2}=\eta \,\, m_{\star}/m,
\end{equation}
where
\begin{equation} \label{eta}
\eta=\frac{3 \int_{0}^{\infty} f(r,u) u du \,\,\,\, \int_{0}^{\infty} 
f(r,u) u^2 du}{f(r,0) \,\,\,\, \int_{0}^{\infty} f(r,u) u^4 du}.
\end{equation}
If $\eta$ is evaluated in the limit $r \rightarrow 0$, it
measures the deviation of the black hole from equipartition of
kinetic energy between it and the stars in the core surrounding
it. When $\eta=1$, there is exact equipartition. We can take various
stellar models and examine what the black hole's stationary state
dynamics would be in each case.

\paragraph*{Maxwellian distribution.}

Suppose the phase space distribution of stars to be given by a
Maxwellian distribution with velocity dispersion $\sigma$:
\begin{equation}
f(r,u) \propto \textrm{exp}(-u^2/2\sigma^2).
\end{equation}
A simple calculation then gives $\eta=1$. This is not surprising: the
analogy in this case is with a Brownian particle immersed in a gas
of molecules with a well-defined temperature \cite{BT};
in the steady state, such a Brownian particle
will be in exact equipartition with the molecules. 

\paragraph*{The King model.}

The simplest model for which the distribution function is a pure
Maxwellian is the ``isothermal sphere'', which is not a very physical
description of actual stellar systems since it has infinite total
mass. A commonly used alternative that resembles the isothermal sphere
at small radii (where stars have large absolute values of energy per
unit mass) is the King model (see \cite{king66}, \cite{BT}):
\begin{equation}
f(E)=\cases{
      \rho_1 (2 \pi \sigma^2)^{-\frac{3}{2}} (e^{-E/\sigma^2}-1)
       &if $E<0$ \cr
       0 &if $E \geq 0$,\cr}
\end{equation}
where $\rho_1$ is independent of $E$. 
In Eq. (\ref{eta}), the upper velocity limit of the integrals is 
now $\sqrt{2 \sigma^2 \xi}$ instead of infinity, where 
$\xi=-\Phi(r)/\sigma^2$.
A calculation reveals that
\begin{equation} \label{etaking}
\eta=\frac{(1-\frac{\xi}{e^{\xi}-1})(\frac{\sqrt{\pi}}{2}
e^{\xi}\,\textrm{Erf}\,(\sqrt{\xi})-\xi^{1/2}-\frac{2}{3}\xi^{3/2})}
{\frac{\sqrt{\pi}}{2}
e^{\xi}\,\textrm{Erf}\,(\sqrt{\xi})-\xi^{1/2}-\frac{2}{3}\xi^{3/2}
-\frac{4}{15}\xi^{5/2}},
\end{equation}
where $\textrm{Erf}\, (z)$ 
is the error function defined as $\textrm{Erf}\,
(z)=2/\sqrt{\pi}\, \int_{0}^{z}\, e^{-t^2}\, dt$. For large values of
$\xi$, 
\begin{displaymath}
\eta \sim 1+(8 \xi^{3/2}/15 \sqrt{\pi}-1)\xi e^{-\xi}+\cdots
O(e^{-2\xi}) \textrm{for} \xi \gg 1.
\end{displaymath}

Fig. 1 shows the relation between $\eta$ and $\xi$. $\eta$ starts at
a value of approximately 1.75 and falls asymptotically to 1 as $\xi$
rises. The King model goes over into the isothermal sphere as $\xi
\rightarrow \infty$. The (mild) deviation from equipartition is because of
the deviation of the distribution function from the Maxwellian
form and, in particular, the existence of an upper velocity limit above 
which the distribution function vanishes. Actual stellar systems are 
well described by
King models with the central potential chosen such that $-\Phi(0)/
\sigma^2$ lies between roughly 3 and 10 (\cite{BT}, \cite{djorgov94},
\cite{kormendy}, \cite{king78}).
As Fig. 1 shows, black
holes at the centers of such systems will be very close to
equipartition with the stars.

\begin{figure}[!t]
\centerline{\psfig{figure=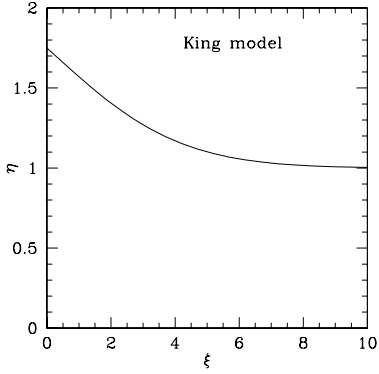,width=2.0in}}
\caption[]{
A plot of $\eta$ as a function of $\xi=-\Phi(r)/\sigma^2$ for a King
model. The Brownian particle at the center of a King stellar model
approaches exact equipartition as $\xi$ becomes large.
}
\label{gpic}
\end{figure}

\paragraph*{Polytropic models.}

Consider next stellar systems described by a polytropic distribution
function which takes the simple form
\begin{equation} \label{polytrope}
f(E)\propto \cases{(-E)^{n} &if $E<0$ \cr
       0 &if $E \geq 0$.\cr}
\end{equation}
Simple calculations give
\begin{equation} \label{npoly}
\eta=(n+5/2)/(n+1) \quad \textrm{for} \quad n>-1.
\end{equation}
Note that $\eta$ for a polytropic model is independent of position
$r$, unlike in the case of the King model. For large values of $n$,
$\eta$ is close to 1, but for small values of $n$, there can be
significant deviations from equipartition.

The potential profile of a polytropic system is a solution of the
Lane-Emden equation \cite{BT}, which can be solved in terms of
elementary functions in only two cases: $n=7/2$ (this case was
considered in \cite{CHL}) and $n=-1/2$. We examine these two cases 
in more detail.

\paragraph*{Polytropic model with $n=7/2$.}

When $n=7/2$, we get the following functional forms for the density
and potential profiles of the stellar system:
\begin{displaymath} 
\rho (r) = \frac{3 M a^2}{4 \pi}\frac{1}{(r^2+a^2)^{5/2}}, \qquad
\Phi (r) = \frac{-G M}{(r^2+a^2)^{1/2}},
\end{displaymath}
where $G$ is the gravitational constant, $M$ is the total mass of the
stellar system in stars, and $a$ is a length
parameter. This is the well-known Plummer model \cite{plum11} which
provides a good fit to some globular clusters. In this case,
$\eta=4/3$, which is mildly different from equipartition. 
$\omega_0^2=k/m=GM/a^3$ for small $r$, and 
\begin{equation} \label{Cplum}
C=(8GM/9a) \gamma m m_{\star},
\end{equation}
\begin{equation} \label{xvplum}
\langle x^2 \rangle = 2a^2 m_{\star}/9m, \qquad
\langle v_x^2 \rangle = 2GMm_{\star}/9am.
\end{equation}

\paragraph*{Polytropic model with $n=-1/2$.}

When $n=-1/2$, we get the following functional forms for the density
and potential profiles of the stellar system:
\begin{displaymath}
\rho (r) = \cases{ 
          M/(4 \pi^2 a^2) \times \sin(r/a)/r &if $r<\pi a$\cr
          0 &if $r \geq \pi a$,\cr}
\end{displaymath}
\begin{displaymath}
\Phi (r) = \cases{
         -GM \sin(r/a)/ \pi r &if $r<\pi a$\cr
         -GM/r+GM/ \pi a &if $r \geq \pi a$.\cr}
\end{displaymath}
In this case, $\eta=4$, which is very different from equipartition. 
$\omega_0^2=k/m=GM/3 \pi a^3$ for small $r$, and 
\begin{equation} \label{Csinr}
C=(8GM/\pi a) \gamma m m_{\star},
\end{equation}
\begin{equation} \label{xvsinr}
\langle x^2 \rangle = 6 a^2 m_{\star}/m,\qquad 
\langle v_x^2 \rangle = 2G M m_{\star}/\pi a m.
\end{equation}

The expressions for $\langle x^2 \rangle$ in Eqs.  (\ref{xvplum}) and
(\ref{xvsinr}) agree with the result of Bahcall \& Wolf \cite{bahcall} to
within factors of order unity.

\paragraph*{Numerical simulations.}

In order to confirm our results, we have performed N-body simulations,
using suitably modified versions of the fourth-order integrators of
Aarseth \cite{aarseth}
developed by Quinlan \cite{quinlan}, for the two polytropic stellar 
systems 
described above. The techniques used are detailed in \cite{CHL} and 
\cite{quinlan}. In both cases, we use units
such that $G=M=1$, and Coulomb forces are softened by $\epsilon=5
\times 10^{-3}$ in order to prevent numerical divergences. Note that
the polytropic initial conditions for the stellar system are not
steady state solutions in the presence of the black hole; the initial
transient response was therefore numerically settled. 

For the case of the Plummer model, we have taken the mass of the black
hole to be $m=0.01$, and the length scale of the potential to be $a=3
\pi/16$. The stellar system is made up of 100,000 stars, each of mass
$m_{\star}= 1 \times 10^{-5}$. The model was integrated for 600 time 
units.
The left half of Fig. 2 shows the binned distributions of one 
component of the
black hole's position and velocity in the stationary state. The solid 
lines show the corresponding bin values as predicted by Eqs. 
(\ref{FPsolstx}) and (\ref{FPsolstv}), under the assumption that in the 
stationary state, ensemble averages will be equal to time averages
over long time spans. 
The agreement with the predictions of the model is evidently good
(these results were described in \cite{CHL}).

The right half of Fig. 2 shows the corresponding plots for 
the $n=-1/2$
polytrope. We have here taken $m=3 \times 10^{-2}$, and $a=3/4
\pi$. The stellar system is made up of 20,000 stars, each of mass 
$m_{\star}= 5 \times 10^{-5}$. The integration time was 900 units. 
Agreement
with the analytical results is again good. We note that the
agreement seen in Fig. 2 would be spoiled if the appropriate value
for $\eta$ were different from that given by our model.

\begin{figure}[!t]
\centerline{\psfig{figure=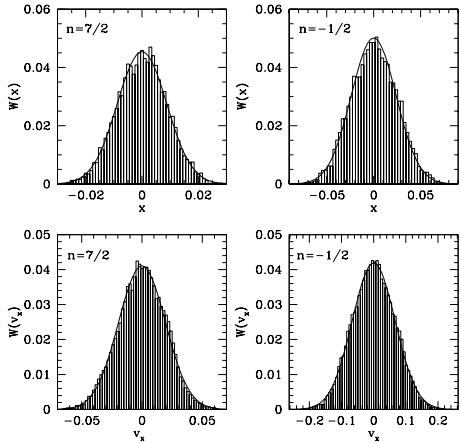,width=2.5in}}
\caption[]{
The panels on the left show the distributions 
of $x$ and $v_x$ from
simulation (bins) and theory (line) for a black
hole of mass $m=0.01$ in a Plummer ($n=7/2$ polytrope)
model made up of 100,000 stars of total mass $M=1$. The panels on 
the right show the
corresponding results for a black hole of mass
$m=0.03$ in a $n=-1/2$ polytrope model of total mass $M=1$
and 20,000 stars.} 
\label{fig:distpic}
\end{figure}

\paragraph*{Summary.}

We conclude that the model outlined in this {\it Letter}
provides a good statistical description of the dynamics of a Brownian
particle -- such as a massive black hole -- in a stellar system. If the 
black hole is inserted at the center of such a system, the transient
effects of its initial position and velocity decay exponentially fast,
and it settles into a stationary state in which the distributions of
its position and velocity become time-independent. 
These distributions are also independent of each 
other and are Gaussian. If the stellar distribution function is
Maxwellian, there is in the stationary state precise equipartition of
kinetic energy between the black hole and the stars. 
The greater the deviation of
the distribution function from a Maxwellian form, the greater is the
deviation from strict equipartition. We have shown several examples of
such cases, in one of which the black hole is very far from
equipartition. The deviation originates from the existence of an upper 
velocity limit for stars beyond which the distribution function is zero.
Note that the stationary state above is not a true equilibrium, since
at very long time-scales (approximately 20 times the relaxation time),
the stellar system will undergo core collapse \cite{BT}. 

The above analytical model applies strictly only to stellar models in
which the density and potential profiles are non-singular at small
radii; in realistic systems, however, the massive black hole would
induce a density cusp consisting of stars bound tightly to it. Since it
carries the cusp with it as it moves around, it is as if a black
hole of a somewhat larger effective mass were moving in a background
of unbound stars whose density profile is flat near the center (and
unaffected by the black hole away from the center provided its mass is
much smaller than the total mass of the stellar system). Since
the restoring force and dynamical friction are provided mainly by the
unbound stars, we would expect our results to apply to real systems
as long as the mass of stars in the cusp is much smaller than the mass
of the black hole; our simulations show that this is indeed so for the 
cases considered here. 

\paragraph*{Acknowledgments.}

We thank G. Quinlan for providing the simulation code.
This work was supported in part by NASA grants NAG 5-7039, 5-7768, 
and by NSF grants AST-9900877, AST-0071019 (for AL).


\begin{references}

\bibitem{chandra43a}
S. Chandrasekhar, Rev. Mod. Phys. {\bf 15}, 1 (1943).

\bibitem{chandra43b}
S. Chandrasekhar, Astrophys. J. {\bf 97}, 255 (1943).

\bibitem{BT}
J. Binney \& S. Tremaine {\it Galactic Dynamics} (Princeton: Princeton
University Press, 1987).

\bibitem{CHL}
P. Chatterjee, L. Hernquist \& A. Loeb, in press, Astrophys. J. (2002). 
(http://xxx.lanl.gov/abs/astro-ph/0107287)

\bibitem{king66}
I. R. King, Astron. J. {\bf 71}, 64 (1966).

\bibitem{djorgov94}
S. G. Djorgovski \& G. Meylan, Astron. J. {\bf 108}, 1292 (1994).

\bibitem{kormendy}
J. Kormendy, Astrophys. J. {\bf 218}, 333 (1978).

\bibitem{king78}
I. R. King, Astrophys. J. {\bf 222}, 1 (1978).

\bibitem{plum11}
H. C. Plummer, Mon. Not. Roy. Astron. Soc. {\bf 71}, 460 (1911).

\bibitem{bahcall}
J. N. Bahcall \& R. A. Wolf, Astrophys. J. {\bf 209}, 214 (1976).

\bibitem{aarseth}
S. J. Aarseth, in {\it Galactic Dynamics and N-Body Simulations},
Ed. G. Contopoulos, N. K. Spyrou \& L. Vlahos (Springer-Verlag:
Berlin, 1994).

\bibitem{quinlan}
G. D. Quinlan \& L. Hernquist, New Astron. {\bf 2}, 533 (1997). 

\end{references}
\end{document}